\documentclass[AMA,STIX1COL]{WileyNJD-v2}
\usepackage{moreverb}
\usepackage{bm}
\usepackage{dsfont}
\usepackage{multirow}
\usepackage{siunitx}
\usepackage{graphicx}
	
\newcommand*\vect[1]{\mathbf{\bm{#1}}}

\newcommand\BibTeX{{\rmfamily B\kern-.05em \textsc{i\kern-.025em b}\kern-.08em
T\kern-.1667em\lower.7ex\hbox{E}\kern-.125emX}}

\articletype{}%

\begin{document}

\title{Misuse of the sign test in narrative synthesis of evidence}

\author[1,2,3]{Stavros Nikolakopoulos}

\authormark{S. Nikolakopoulos}

\address[1]{\orgdiv{Evidence Synthesis Methods Team}, \orgname{Department of Primary Education, University of Ioannina}}

\address[2]{\orgdiv{Department of Statistics}, \orgname{Athens University of Economics and Business}}

\address[3]{\orgdiv{Department of Biostatistics}, \orgname{University Medical Center Utrecht}}

\corres{*\email{snikolakopoulos@uoi.gr}}


\abstract[Abstract]{In narrative synthesis of evidence, it can be the case that the only quantitative measures available concerning the efficacy of an intervention is the direction of the effect, i.e. whether it is positive or negative. In such situations, the sign test has been proposed in the literature and in recent Cochrane guidelines as a way to test whether the proportion of positive effects is favourable. I argue that the sign test is inappropriate in this context as the data are not generated according to the Binomial distribution it employs. I demonstrate possible consequences for both hypothesis testing and estimation via hypothetical examples.}


\maketitle

\subsection*{Acknowledgments} 
This is the pre-peer reviewed version of the following article: \textit{S Nikolakopoulos, Misuse of the sign test in narrative synthesis of evidence, \textbf{Res Synth Methods} 2020 Sep;11(5):714-719. doi: 10.1002/jrsm.1427} which has been published in final form at https://onlinelibrary.wiley.com/doi/abs/10.1002/jrsm.1427. This article may be used for non-commercial purposes in accordance with Wiley Terms and Conditions for Use of Self-Archived Versions.

The author has received financial support from the COMPAR-EU project (Comparing effectiveness of self-management interventions in 4 high priority chronic diseases in Europe), H2020 - Grant agreement 754936 as well as from the DRASI-II funding scheme, Athens University of Economics and Business, project "Bayesian Variable Selection for Network Meta Analysis". 


\section{Introduction}

In situations where essential ingredients required for a meta-analysis to be conducted are not available, some form of "narrative synthesis" is undertaken. Narrative synthesis is a rather common approach to evidence synthesis. For example, recent results indicate that it is employed in more than half of the systematic reviews evaluating public health interventions \cite{campbell2019}. The approach followed for synthesizing the evidence in narrative reviews, depends on the amount of detail present in the the available data. The focus of this note is on the case where the only possible summary of effects is that on the \textit{direction} of the effect, i.e. whether there is an overall positive or negative effect of an intervention. This could be the case when only direction of effect is reported, or there is inconsistency in the effect measures or data reported across studies \cite{mckenzie2019}. 

Common approaches for summarizing results for effect direction are graphical representations of the available data, such as the harvest plot \cite{ogilvie2008} or the effect direction plot \cite{thomson2013edp}. The main objective of such plots is to visually present findings on the direction of, possibly different, outcomes. Such outcomes might be conceptually similar but measured by different endpoints. The counting of outcomes that exhibit a positive outcome (sometimes termed \textit{vote counting}) may be based on the statistical significance of each outcome. This practice has been acknowledged to be misleading and is considered not acceptable, since variable power across studies can lead to misleading conclusions\cite{mckenzie2019}.

As an alternative, the latest Cochrane Handbook for Systematic Reviews of Interventions \cite{mckenzie2019}, suggests vote counting to be conducted on the basis of the direction of the effect of each outcome, i.e. whether it is positive or negative and not whether it is statistically significant or not. It is furthermore suggested that one could conduct a sign test, i.e. a test based on the Binomial distribution with the null hypothesis being that $\pi=0.5$ to answer the question ‘is there any evidence of an effect?’. It is there mentioned that "this is equivalent to testing if the true proportion of effects favouring the intervention (or comparator) is equal to 0.5 ... An estimate of the proportion of effects favouring the intervention can be calculated (p = u/n, where u = number of effects favouring the intervention, and n = number of studies) along with a confidence interval (e.g. using the Wilson or Jeffreys interval methods)". 

The suggestion of using a two-sided sign test for testing the above hypothesis has been long-standing in the literature \cite{hedges1980,bushman1994,borenstein2011,mckenzie2014}. In what follows, I argue that the binomial distribution is inappropriate for modelling the proportion of favourable effects. As a consequence, the two-sided sign test is unsuitable for testing the above hypothesis and, if evaluation of statistical significance is deemed necessary, the one-sided version may be used and interpreted with great caution. Furthermore, interval estimation might be misleading. The purpose of this brief note is expository and the focus will be on an example rather than on lengthy theoretical arguments.

\section{The sign test}
\subsection{Definition}

Let $X$ be a random variable following a binomial distribution with parameters $n$, the number of experiments and $\pi$, the probability of success in each experiment and thus $X\sim Bin (n,\pi)$. A fundamental assumption of the Binomial distribution is that $\pi$ is fixed and constant across all the $n$ experiments. The sign test is merely a hypothesis test for $H_0:\pi=0.5$. For $x$ observed successes in $n$ experiments, the one sided $p$-value towards $H_+:\pi>0.5$ is equal to $Pr(X\geq x|\pi=0.5)$, while it is $Pr(X\leq x|\pi=0.5)$ towards $H_-:\pi<0.5$. The two sided $p$-value towards  $H_{\neq}:\pi \neq 0.5$ is defined as $2\min \left\{Pr(X\geq x|\pi=0.5),Pr(X\leq x|\pi=0.5) \right\}$ \cite{conover1980}. It is an exact test, in the sense that it does not rely on asymptotic approximations, but it rather employs the exact distribution of the underlying data. For fixed $n$ and significance level $\alpha$ (one sided, thus corresponding to 2$\alpha$ two-sided)  a critical value $c_{n,\alpha}$ can be found such that $Pr(X \geq c^+_{n,\alpha}  |\pi=0.5) = Pr(X\leq c^-_{n,\alpha}  |\pi=0.5) \leq \alpha$. The equality is a result of the symmetry of the Binomial distribution when $\pi=0.5$ and as a consequence $c^+_{n,\alpha}=n-c^-_{n,\alpha}$. The events $R_+= \{X\geq c^+_{n,\alpha} \}, R_-= \{X\leq c^-_{n,\alpha}\}$ denote rejection in favour of $H_+$ and $H_-$ respectively.


\subsection{Interpretation for vote counting in evidence synthesis}\label{int}

When used as suggested, the Binomial distribution is employed to model the probability of a study showing a positive effect. Without loss of generality, I assume larger values on some outcome $Y$ indicate a positive treatment effect, and thus the term "positive effect" in the context discussed here is meant such that $E(Y)>0$. The data observed are $\hat{Y}_i$ the treatment effect estimates in study $i \in \{1,\dots,n\}$ and $X_i=\mathds{1}_{\{\hat{Y}_i>0\}}$ where $\mathds{1}_{\{ \cdot \}}$ the indicator function. Thus, by employing the sign test and interval estimation methods, one assumes $X=\sum X_i \sim Bin (n,\pi)$. $\hat{Y}_i$ could be any treatment effect measure such as the difference in means for continuous data or the $\log OR$ for binary data. The probability $\pi$ is thus meant to model $Pr(\hat{Y}_i>0)$, equal and fixed across studies and also interpreted as "the proportion of effects favouring the intervention". Note that $Pr(\hat{Y}_i>0)=Pr(P_i<0.5)$, where $P_i$ the $p$-value associated with (standardized) $\hat{Y}_i$. 

\section{Issues with the sign test in evidence synthesis}

In what follows I assume that each study $i$ is testing the mean of a single group, to avoid further notational complexity. The extension is straightforward to the two-sample case, the standard case when dealing with clinical trials' synthesis. If the (individual, i.e. patient) data in study $i$ are generated from a normal distribution with mean $\mu$, variance $\sigma^2$ and $\delta=\mu/\sigma$, then the cumulative distribution function of the $p$-value for testing $H_0:\mu=0$ vs $H_1:\mu>0$ is \cite{hung1997}	

\[
F(P_i)=Pr(P_i\leq p)=1-\Phi (Z_p - \sqrt{N_i}\delta_i)
\]
where $Z_p$ the (1-$p$)th percentile of the standard normal distribution, $N_i$ the sample size of study $i$ and $\delta_i$ the true standardized treatment effect in study $i$. As a result, for the Binomial distribution implied in Section \ref{int}, $\pi=Pr(P\leq 0.5)=\Phi(\sqrt{N_i}\delta_i)$ which is fixed and equal to 0.5 iff $\delta_i=0,$ $\forall i$. For any deviation from $H_0$ and assuming that not all studies have identical sample and effect sizes, the distribution of $X$ is not Binomial, but a Poisson-Binomial \citep{hong2013} $PoisBin(\vect{\pi}_n)$. The vector of probabilities $\vect{\pi}_n=(\pi_1,\dots,\pi_n)$ depends on each study's and outcome's sample size and true effect size via $\pi_i=\Phi(\sqrt{N_i}\delta_i)$. Thus, the null hypothesis of the sign test in this case does not define only the value of the parameter tested but also the distribution of the data-generating model. Possible implications of this approach are outlined below.  

\subsection{Implications for hypothesis testing}
It has been suggested that two-sided testing may be conducted for vote counting in the context discussed here. The null hypothesis dictates that all $\delta_i$'s are 0 which translates to $H_0:\pi=0.5$. The alternative is described as "the proportion of studies with a positive outcome is larger than 0.5". The only way to formulate such an alternative within the Binomial distribution context is $\pi>0.5$, which translates to all $\delta_i$'s and $N_i$'s being identical, a rather unrealistic assumption - especially given the application of the sign test for different outcomes. Such an alternative could be described by the Poisson-Binomial model as 

\begin{equation}\label{eq:1}
H_1: \sum_i{\mathds{1}_{\{\pi_i>0.5\}}}>\frac{n}{2}
\end{equation}
which is neither obvious how it should be tested, nor one has available data to estimate all $\pi_i$'s in a given sample of studies. 

Furthermore, for the Binomial distribution and $H_0:\pi=0.5$, when using either one or two-sided p-values any deviation towards $H_+$ results to $Pr(R_+)>Pr(R_-)$. The reverse is true for deviations towards $H_-$. That is, the power function is monotonically increasing in $\pi$ for the respective alternative, a desirable characteristic for a test statistic. This is not the case when the data are actually generated from a Poisson-Binomial distribution and the Binomial distribution is used for testing $H_0$. There can exist configurations of $H_1$ as described in \ref{eq:1}, where $Pr(R_+)<Pr(R_-)$. In other words, there can be situations where the number of positive effect sizes is larger than that of the negative ones, and the probability of rejecting $H_0$ towards $H_-$ (and thus claiming a negative overall effect) is larger than rejecting towards $H_+$. An example is used below to illustrate. 

\subsection{Implications for estimation}
For hypothesis testing, parameters are set to a fixed value assumed under $H_0$ and the distribution of the test statistic is utilized. For estimation purposes there is no underlying hypothesis concerning the values of relevant parameters and the (usually symmetric) distribution of an estimator emerges, based on the distribution of the underlying data and/or asymptotic approximations. As described above, the Binomial distribution is clearly not suitable for describing the proportion of positive effect sizes in the general case (not under $H_0$) and thus confidence intervals based on the normal approximation of the assumed Binomial distribution (Wilson interval\cite{brown2001} ) or on Bayesian formulation based on the assumption of a single $\pi$ (Jeffreys interval\cite{brown2001} ) are inappropriate. Here I will use these two intervals suggested in the handbook to demonstrate this realization. 

\section{Example}
As a basis for demonstration, I use the example presented in the Cochrane handbook \cite{mckenzie2019}, Chapter 12.4.2.3. Following the handbook's description closely, a review examined the effects of midwife-led continuity models versus other models of care for childbearing women. The outcome of interest is maternal satisfaction. There were differences in the concepts measured, the measurement period, and the measurement tools. In scenario 4 of the example, there is minimal reporting of the data, and the type of effect measure varies across the studies (e.g. mean difference, odds ratio). The final dataset included 12 studies, in 10 of which the respective outcome favoured midwife-led models of care. 

The theoretical data generating model is $Bin(12,\pi)$ for the sign test. As discussed above, the actual data generating model is $PoisBin(\vect{\pi}_{12})$ where $\pi_i=\Phi(\sqrt{N_i}\delta_i)$ which is only equal to $Bin(12,\pi)$ under $H_0$ and thus $\pi_i=0.5$ $\forall i$. For ease of demonstration, I will employ hypothetical configurations of $\vect{\pi}_{12}$ where $K$ of the 12 $\pi_i$'s are equal to a large value, $\pi_{L}>0.5$ corresponding to studies with positive true effect sizes and $n-K$ are equal to a small value, $\pi_{S}<0.5$ corresponding to studies where $\delta<0$. 

For hypothesis testing and one-sided $\alpha/2=0.025$ and $n$=12, $c^-_{12,0.025}=2$. Therefore, any systematic review with 12 studies that finds a positive effect on more than nine or less than three studies, should conclude a positive or negative effect of the treatment, respectively, if the two-sided sign test is employed with $\alpha=0.05$. The parameter of interest is $\pi_+=K/n$, the proportion of studies with a positive effect size. If $\pi_+>0.5$ it is desirable that $Pr(R_+)>Pr(R_-)$ and the coverage probability of a $(1-\alpha)\%$ confidence interval should be around $(1-\alpha)\%$. The exact significance level for each of the sides for the sign test is $Pr(R_+|H_0)=Pr(R_-|H_0)=0.019$.

\begin{figure}[!t]
\begin{center}
\makebox[0cm]{ \includegraphics[width=0.7\textwidth]{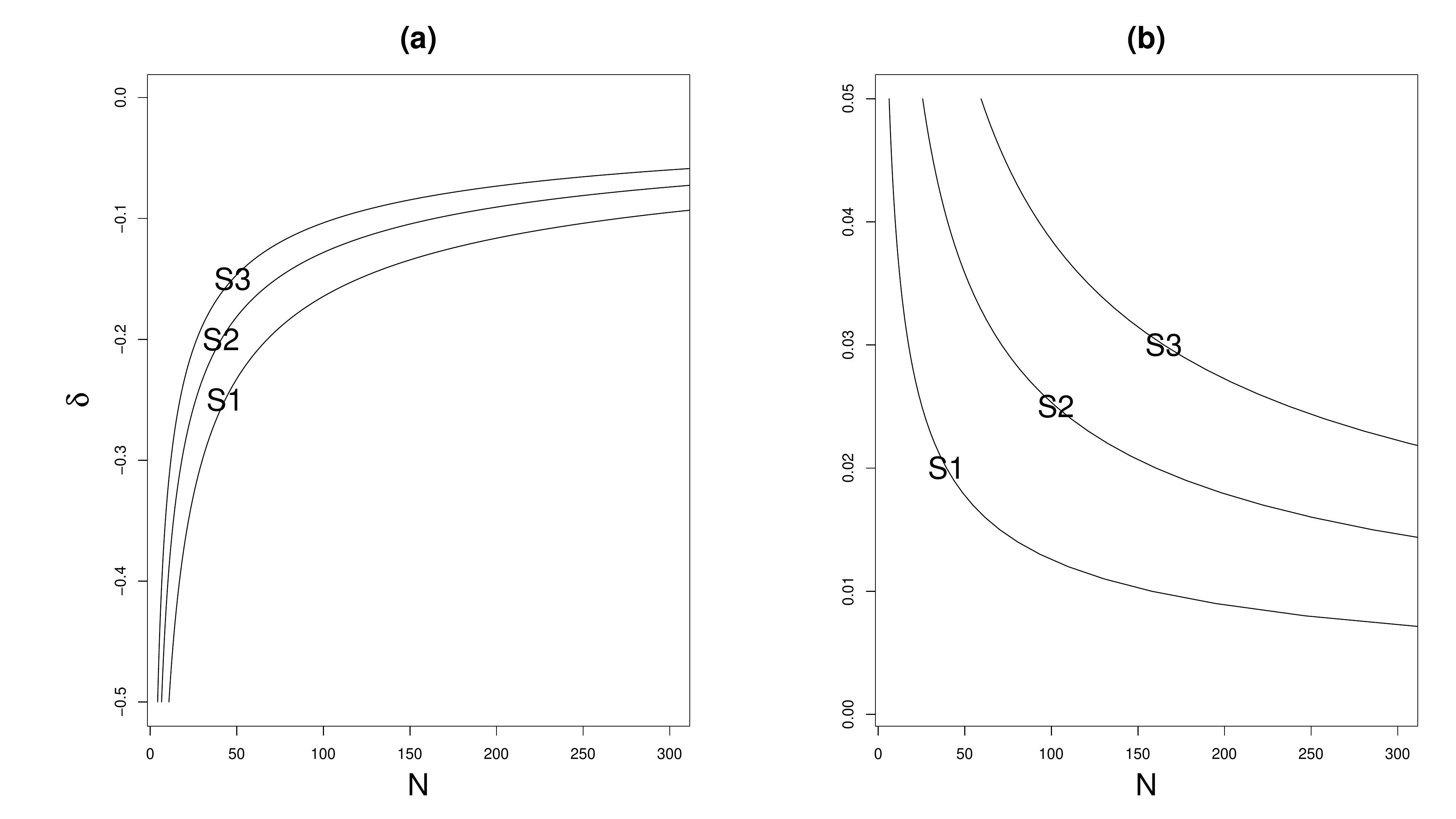}}
\end{center}
\caption{Possible combinations of effect size and sample size per study so that $\pi_S$ (a) and $\pi_L$ (b) are as described in the respective scenario S1,S2 and S3}
\label{fig:1}
\end{figure}

I consider three scenarios. Scenario 1 assumes that $\pi_{S}=0.05$ and $\pi_{L}=0.55$. For Scenario 2  $\pi_{S}=0.1$ and $\pi_{L}=0.6$ while in the last scenario $\pi_{S}=0.15$ and $\pi_{L}=0.65$. Possible combinations of $\delta$ and $N$ for each scenario are presented in Figure \ref{fig:1}. The described scenarios are merely for demonstration as similar test characteristics could be the result of countless configurations of $\vect{\pi}_{12}$, which in turn is a function of infinite combinations of $\delta_i$'s and $N_i$'s.

Table \ref{tab:1} describes the results of this exercise. It shows scenarios where $Pr(R_-)> Pr(R_+)$, while $\pi_+>0.5$ and $H_1$ as described in eq. \ref{eq:1} holds. It depicts situations where the proportion of positive effects is favourable, but the Binomial test has a higher probability of favouring negative effects. This is a demonstration of why two-sided testing is inappropriate in such cases and how even one-sided testing would significantly lack power ($Pr(R_+)$ very close to 0, and considerably smaller than $\alpha$). It is also shown (via a simulation exercise where data are generated according to the Poisson-Binomial distribution and $\hat{\pi}_+=K/n$) how confidence intervals based on the binomial distribution fail to estimate the required proportion adequately. In the most extreme scenario 1, even when only 2 out of 12 studies have a negative effect, the probability of rejecting towards $H_-$ is five times larger than $Pr(R_+)$. 

\begin{table}[t!]
\centering
\begin{tabular}{cccccrrrr}
&&&&&&&\multicolumn{2}{c}{95\% CI Coverage} \\
\cline{8-9}
Scenario&	$\pi_{S}$&$\pi_{L}$&	$K$&$\pi_+$& $Pr(R_-)$& $Pr(R_+)$&	Jeffreys& Wilson \\
\hline
\multirow{4}{*}{1}&	\multirow{4}{*}{0.05}&	\multirow{4}{*}{0.55}&	7&	0.58&	0.126&	<0.001&	0.666&0.662 \\
&	&	&	8	&0.67&	0.075&	<0.001&	0.530&0.528 \\
&	&	&	9	&0.75&	0.044&	0.001&	0.397&0.180 \\
&	&	&	10	&0.83&	0.025&	 0.005&	0.115&0.117 \\
\\
\multirow{2}{*}{2}&	\multirow{2}{*}{0.10}&	\multirow{2}{*}{0.60}&	7&	0.58&	0.063&	<0.001&	0.797&0.798 \\
&	&	&	8	&0.67&	0.035&	 0.001&	0.680&0.678 \\
\\
3&	0.15&	0.65&	7&	0.58&	0.028&	 0.002&	0.889& 0.888\\
\hline
\end{tabular}

\caption{Consequences of using the sign test for evaluating the proportion of studies with positive effects, in configurations where the proportion is favourable $(\pi_+>0.5)$ but highly heterogeneous. Scenarios are described in detail in the text. $\pi_+$ denotes the proportion of favourable effects. $R_-$ denotes the rejection region of the sign text in the unfavourable side, while $Pr(R_-)$ denotes the probability of falling in this region based on the data -generating model of the Poisson-Binomial distribution. Coverage of the $95\%$ Jeffreys and Wilson intervals for $\pi_+$ based on $10^4$ simulations is also presented.}
\label{tab:1}
\end{table}

\section{Conclusions}
I argue that the sign test is inappropriate for testing the hypothesis that the proportion of effects favouring an intervention is larger than 50$\%$ in a narrative synthesis that only effect direction is considered. It is clearly not appropriate for estimating this proportion since the data-generation mechanism in the general case (Poisson-Binomial distribution) is not in accordance with the estimation assumptions (Binomial distribution). From a hypothesis-testing perspective, it can lead to awkward situations of $Pr(R_-)> Pr(R_+)$ when the parameter of interest is in the opposite direction. This is undesirable for two-sided testing and can lead to very low power (even $<\alpha$) for one-sided testing. 

The scenarios I examined are clearly extreme (as depicted in Figure \ref{fig:1}). In such cases one could argue what the actual average treatment effect is and whether we would like to find such an effect. However, since it is suggested that the sign test be used when combining different outcomes, it is not unrealistic to consider cases where the effect is positive in some and negative in others. This could also be the result of observed heterogeneity on very similar outcomes. In such cases, i.e. for a mean treatment effect distributed as $N(\theta,\tau^2)$, the cdf of the $p$-value is still tractable \cite{hung1997} and similar calculations could be performed. 

The literature on the abuse of statistical significance and $p$-values is vast \cite{sterne2001} and still very much in fashion \cite{wasserstein2019}. Perhaps narrative reviews of the direction of the effect is one of the cases where researchers could be allowed to refrain from statements concerning statistical significance and employ the very informative visual representations suggested in the literature \cite{ogilvie2008,thomson2013edp}.

\bibliography{SignTest}

\end{document}